# The Coriolis Effect Apparently Described in Giovanni Battista Riccioli's Arguments Against the Motion of the Earth: an English rendition of *Almagestum Novum* Part II, Book 9, Section 4, Chapter 21, pages 425, 426-7.

Christopher M. Graney

Jefferson Community & Technical College
1000 Community College Drive
Louisville KY 40272 (USA)
christopher. graney@kctcs.edu

In his encyclopedic work on astronomy, the 1651 *Almagestum Novum*, the Italian Jesuit Giovanni Battista Riccioli (1598-1671) argued against the movement of the Earth on the grounds that (among other things), if the Earth rotated, that rotation should produce a deflection in the trajectories of projectiles. This argument appears to be an early description of the Coriolis effect.

Within his 1651 *Almagestum Novum*[1] the Italian Jesuit astronomer Giovanni Battista Riccioli (1598-1671) included, among many other things, 77 arguments against the Copernican motion of the Earth. Prominent among these were several arguments based on the Earth being a rotating frame of reference. According to Riccioli, the Earth's rotation should produce deflections in falling bodies and projectiles.[2] Such effects today are discussed in physics textbooks as manifestations of the Coriolis Force.[3]

What follows is a rendition in English of two of these "Coriolis" arguments (*Almagestum Novum* Part II, Book 9, Section 4, Chapter 21, pages 425, 426-7), which are arguments number 17 and 19 of Riccioli's 77.[4] It is not a close translation, for in places we (I thank Christina Graney, without whose efforts in translating Riccioli's Latin this work would not exist) significantly condense or re-arrange what Riccioli says so as to produce a rendition both palatable to the modern reader and true to Riccioli's meaning; Riccioli's Latin sentences in these arguments have lengths of as much as 159 words. As an example – we condense Riccioli's writing on how a cannon ball "might barely hit target N, as if by swift and very oblique contact, which might more merit the name of abrasion or burning rather than striking and wounding"[5], to say the ball might "graze" the target.

Note – we delineate Riccioli's paragraphs by spaces. Often we divide these into smaller paragraphs, which we indicate by indented lines. The numbering system is Riccioli's.

---

1  G. B. Riccioli, *Almagestum Novum* (Bologna, 1651) <http://www.e-rara.ch/zut/content/pageview/140188>, henceforth *AN*. All page numbers cited here are from Part II of the *AN*.

2  Christopher M. Graney, "Giovanni Battista Riccioli's Seventy-Seven Arguments Against the Motion of the Earth: An English Rendition of *Almagestum Novum* Part II, Book 9, Section 4, Chapter 34, Pages 472-7", arXiv:1011.3778v1 <http://arxiv.org/abs/1011.3778> (2010).

3  Jerry B. Marion, *Classical Dynamics of Particles and Systems* (Academic Press/Harcourt Brace Jovanovich: Orlando, Florida, 1970), 343-56.

4  Graney, "...Riccioli's Seventy-Seven Arguments..."

5  "sed veluti fugaci ac valde obliquo contactu, qui mereretur potius nomen attritionis aut confrictionis, quam ictus & vulneris, vix feriret scopum N [*AN,* II, 427, col. 1]"

**VI.** *The argument regarding balls from a Cannon fired near the poles and on different Parallels of latitude.*[6]

VIII.  Tycho also argues that if the cannon experiment were performed at the poles of the Earth, where the ground speed produced by the diurnal motion is diminished, then the result of the experiment would be the same regardless of toward which part of the horizon the cannon was fired.  However, if the experiment were performed near the equator, where the ground speed is greatest, the result would be different when the ball is hurled East or West, than when hurled North or South.

The form of the argument is thus:  *If Earth is moved with diurnal motion, a ball fired from a cannon in a consistent manner would pass through a different trajectory when hurled near the poles or toward the poles, than when hurled along the parallels nearer to the Equator, or when hurled into the South or North.  But this is contrary to experience.  Therefore, Earth is not moved by diurnal motion.*

If Tycho is to be believed, experiments have shown this to be correct.  Moreover, if a ball is fired along a Meridian toward the pole (rather than toward the East or West), diurnal motion will cause the ball to be carried off [i.e. the trajectory of the ball is deflected], all things being equal: for on parallels nearer the poles, the ground moves more slowly, whereas on parallels nearer the equator, the ground moves more rapidly.[7]

---

6   *AN*, II, 425, col. 2, from the center of the column to the bottom of the page.

7   For key passages which we believe describe the Coriolis effect, we provide the reader with the original Latin, and in some cases an alternate, more literal, translation.  The Latin here reads "…quia si globus exploderetur versus polos per planum euisdem Meridiani, minor illi diuersitas a motu diurno inferretur, quam si modo versus Ortum, modo versus Occasum; Si vero in parallelis polo propioribus, tardius cum terra, si in propioribus Æquatori celerius cum terra ferretur ille globus, cœteris, vt supponitur, paribus [*AN*, II, 425, col. 2.]."

The Copernican response to this argument is to deny it, or to concede it but claim that the differences in trajectory fall below our ability to measure. But in fact the argument is strong, and this response is not.

**VIII.** *The argument against the Diurnal and Annual Motion of the Earth devised by Father Francisco Maria Grimaldi[8] based on a Cannon ball fired into the North, and into the East and West.[9]*

X. Until we can observe the trajectories which bodies pass through, the arguments against the motion of the Earth based on those trajectories will lack strength. However, if we consider where the bodies strike the ground and their impetuses at impact, we will have stronger weapons with which to argue against the Earth's motion.[10] What follows is an argument based on cannon ball strikes.

Suppose that a very large cannon ball, weighing 60 or 80 pounds, traverses 250 paces in 2 human pulsebeats, or 2 seconds. Those skilled in using artillery consider this reasonable.

Refer to Figure 1. A cannon whose mouth is at A is aimed at a target at B which is 250 paces East of A. Thus a cannon ball fired from A arrives at B in 2 seconds if the Earth does not move. If Earth has a diurnal rotation, then in 2 seconds of time, both cannon and target move through 30 seconds of arc, or 752 paces at the equator.[11] Thus both the mouth of the cannon and the target traverse 752 paces – the mouth from A to C, and the target from B to D. So the mouth and target remain separated by 250 paces. Ball I fired from A hits the target at D.

---

8  Grimaldi (1618-1663) was a close associate of Riccioli.

9  *AN*, II, 426, col. 2, center of the column to 427, col. 2, lower part of the column. A discussion of part of this argument can be found in Edward Grant, "In Defense of the Earth's Centrality and Immobility: Scholastic Reaction to Copernicanism in the Seventeenth Century", *Transactions of the American Philosophical Society*, New Series, Vol. 74, No. 4 (1984), starting on page 48.

10 Here Riccioli directs the reader to see chapter 19.

11 Riccioli directs the reader to a table in chapter 19, number 13, which states the circumference of the Earth as being 32,512,000 Roman paces.

The cannon is now aimed north to target E. The distance AE is equal to AB, 250 paces. The cannon is allowed to cool. Then, keeping everything else the same – the same ball, the same quantity and quality of gunpowder, the same inclination or elevation of the cannon, the same air conditions – if the ball is fired from A, it will travel straight to E, if Earth is at rest.

But if Earth and all bodies related to it move by diurnal motion, during the two seconds in which the cannon ball is traveling from the cannon to the target, the cannon is transported 752 paces, from AQ to CR. The target will be transported to N, and the cannon ball to F, where it hits. It appears that the ball traverses 250 paces, while in fact, it travels farther along a trajectory AKF – whose chord AHL is 825[12] paces as calculated by trigonometry. Angle AFC is 70 degrees, 35 minutes[13] and equal to angle NFM.

The cannon ball is going to hit the northern target N more weakly than it will hit the eastern target D. So the ball should be observably less effective in breaking a city wall or smacking a second ball, at N than at D. There are two concurrent causes of the weakened impetus of the ball sent to the north.

*First*, consider the impetus the ball has as it leaves the mouth of the cannon at A. It is such that if Earth is immobile, the ball hits B or E, after traveling through a straight line (AB or AE) distance of 250 paces. But if the Earth is moving, both the ground and the ball are moving toward the east. Thus while the ball moves toward E, owing to the prevailing diurnal motion, it moves not in the straight line AE, but in a curve AKF toward F. (This happens because the diurnal motion decreases as the ball travels farther from the cannon to the northern target, and so the ball travels beyond the line AHF which it would follow if the diurnal motion were uniform.)[14]

---

12  This value should be 792.5; calculation errors are not uncommon in the *Almagestum Novum* (see C. M. Graney, "The Telescope Against Copernicus: Star Observations by Riccioli Supporting a Geocentric Universe", *Journal for the History of Astronomy*, vol. 41 (2010), 453-67).

13  71 degrees, 37 minutes.

The ball's impetus is necessarily lessened by the diurnal motion as the ball is compelled to follow the longer path AKF rather than the short path AE. The reason for this lessening, perhaps significantly, of impetus can be understood as follows: Consider a ball dropped perpendicularly through a pipe that is moving horizontally, so that the ball, after leaving the pipe, strikes against a hand below the lower end of the pipe. Now consider the same situation, but with the pipe stationary. In the first case the impetus against the hand is much less.[15]

*Second*, from C, looking along FN, the cannon ball may be seen to hit target N at point F. But it might hit F with an oblique blow along LM. Owing to the diurnal motion, the impetus will be deflected from straight lines AE or FN to line AHL. More correctly, the impetus is deflected along curve AKF so that the angle it makes with CF is more than the angle AFC, which we calculated to be 70 degrees, 35 minutes. So the ball does not collide with the target along LGM, but rather its impetus diverges from LGM to the extent that it might not impinge directly into the target at all, but merely graze it or miss it.

---

14 More literally: "Whereas because both earth and the ball are assumed to be carried off toward the East, and while the ball advances toward E, by the prevailing diurnal motion it departs and is turned away from the straight [line] AE, and it is drawn through the curve AKF, toward F, (because in the beginning this motion is faster, and the ball is brought beyond the straight [line] AHF, which it might trace out if the motion be uniform)..."; "At quia & tellus & globus transferri ponuntur Orientem versus, & dum globus nititur versus E, prævalente diurno motu deuiat ac detorquetur a recta AE, trahiturque per curuam AKF, versus F, (quia in principio motus hic velocior est, & globus fertur vltra rectam AHF, quam describeret si motus esse vniformis)...[*AN*, II, 427, col. 1.]".

15 The statement that the impetus at the point of impact is reduced significantly by the diurnal motion seems contrary to the idea of impetus as described by J. Buridan. Buridan states that impetus is proportional to the quantity of matter present in an object; is proportional to the speed of the object; is directional; only decreases if there are corrupting influences present, such as air [J. Buridan, "The Impetus Theory of Projectile Motion", translated from Latin into English by Marshall Clagett, as found in *A Source Book in Medieval Science* by Edward Grant, editor (Cambridge Massachusetts: Harvard University Press, 1974), 275-277]. Riccioli and Grimaldi recognize the relative motion of cannon and ball as seen from their discussion of the ball fired due East. We do not understand why they think that the ball arrives at generally the same location after the same interval of time, and thus has speed unchanged relative to the target whether or not Earth has diurnal rotation, but has perhaps significantly altered (reduced) impetus. Furthermore, this idea that impetus is significantly reduced is contrary to statements they make in an upcoming passage.

Suppose another target is positioned to the side of N on the right toward the East, such as at G. Even if the cannon is not aimed toward it, the ball might hit it with very great impetus,[16] with the impact being in the direction of the impetus. It will be capable of breaking through a city wall and causing great ruin.[17] And if targets N and G were equal in all other respects and were such that they could be moved by the impact of a cannon ball, target G would be driven much farther, traveling through GM, while target N would be driven less far through FN. For we see in the game of billiards that if one ball strikes another tangentially, it delivers a much weaker blow than if it strikes directly. Likewise in the game of tennis, a racket delivers a much weaker blow to the ball when swung with a chopping motion than when swung so as to strike the ball perpendicularly.

None of the above examples of what should happen if the Earth moves are in accord with what we see. Therefore, the Earth does not move with diurnal, much less annual, motion.

Now comes the formal argument:

*If Earth moves by diurnal or annual motion the impact of a cannon ball fired into the North or South should be less effective than one fired into the East or West. We do not see this*

---

16  More literally: "and if to the side of N itself, another target might be at G, located on the right toward the Greek [East] wind, even if the cannon and ball might not be directed into that [target]; still the ball might hit that by very great impetus"; "& si ad latus ipsius N, esset scopus alter in G, dextrorsum versus ventum Græcum collocatum, etiam si in illum non esset directa Bombarda & globus; illum tamen multo maiori impetu feriret globus"

17  This statement concerning impetus seems to contradict the earlier statement that impetus of the ball would be lessened.

*consequent effect, therefore the Earth must not move.* This holds true for other projectiles as well. Skilled artillerymen can aim so well as to place a shot down the mouth of an enemy cannon. So, they can aim well enough that this effect caused by the diurnal motion should be noticeable and should have been noticed.

    The Copernicans might answer that the cannon, the ball, and the target are all transported together with the same velocity. So, while the ball takes a slanted path from the cannon to the target, its impact at the target is, nevertheless, direct. Suppose, on a ship traveling toward the East, someone looks through a narrow pipe at a burning candle (also on the ship) to the North of the person; the view will be the same as if the ship was not moving. Likewise if a billiard table is on the ship and one ball is struck so as to hit a second ball to its North, even though the ship is moving toward the East, the ball will hit directly. And lastly, if a sailor swings a hammer toward a mast at his north, he will drive in a nail even though he and the mast are moving toward the east. The ball and the hammer, despite the motion of the ship and the slanting way they move because of the two motions mixing together, will hit the desired locations.

    This response works insofar as the impacts in these examples remain square; still, the impetus of the impacts will be weakened by the transverse motion of the ship that carries the ball or hammer.[18] We have made many experiments that convince us of this. Regardless, in the example of the billiards, the hammer, and the ship, the distances involved are so small that these effects are not noticeable, and in the case of the hammer, the arm would compensate for them anyway.

---

18  Here again we see the apparent claim that impetus is altered even if speeds and directions are not.

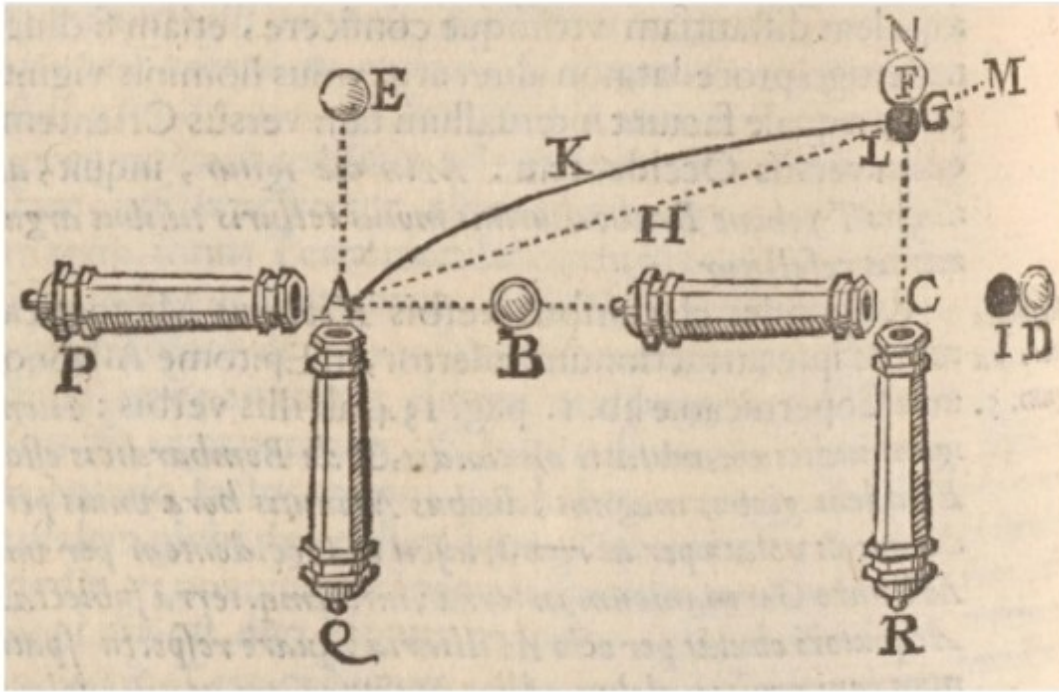

Figure 1a: Illustration from the *Almagestum Novum*. A cannon with mouth at A is fired toward Eastern target B and Northern target E, both equally distant from A. While the cannon ball (I and F) is in flight, the diurnal rotation of the Earth carries the cannon mouth to C, the Eastern target to D, and the Northern target to N.

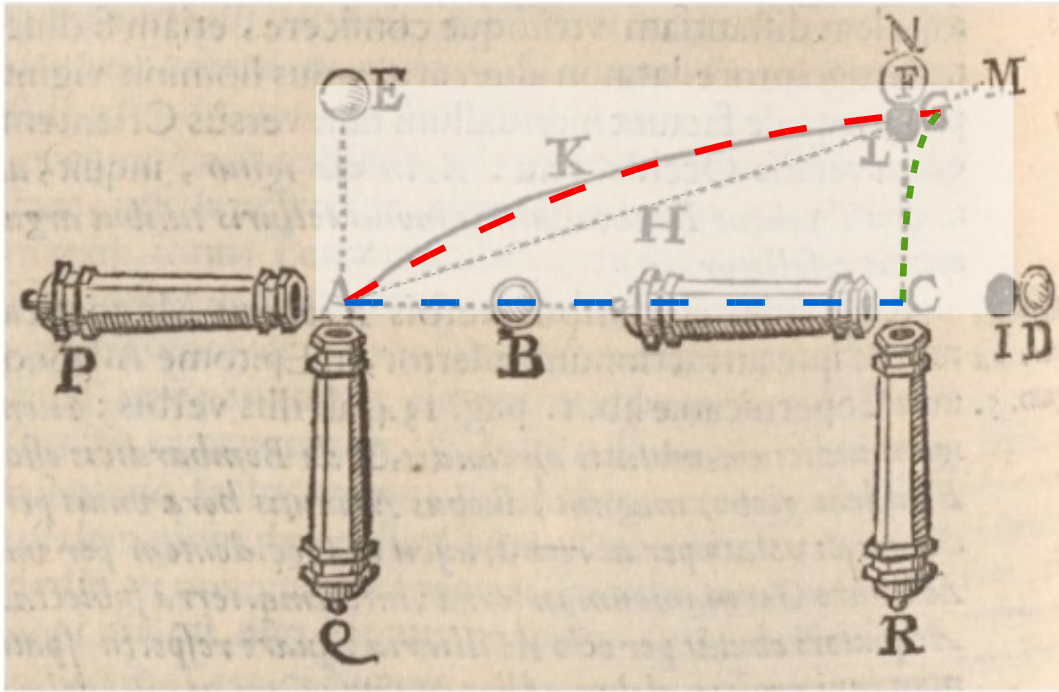

Figure 1b: Plots of simple Coriolis force calculations, superimposed on the illustration from the *Almagestum Novum*. Colored dashed lines show paths for a cannon in a rotating frame of reference, firing a ball toward the axis of rotation. Blue line is the path of the cannon's mouth; red line is the path of the ball; green line is the path of the ball as seen from the cannon, drawn from position C. Up (North) is toward the axis of rotation. The scale of the plot has been fit to the illustration, and the distance to the axis of rotation has been adjusted so as to yield agreement between the red line and curve AKF. Nevertheless, note the rightward (Eastward) deflection of the ball as seen from the cannon, away from line of fire CF and into G, as described by Riccioli and Grimaldi.